\documentclass{emulateapj}

\usepackage{graphics,graphicx}
\usepackage{natbib}
\usepackage{epstopdf}
\def\apjl{\rm ApJL}
\def\apjs{\rm ApJS}

\def\aap{\rm AAP}

\newcommand{\gtorder}{\mathrel{\raise.3ex\hbox{$>$}\mkern-14mu
            \lower0.6ex\hbox{$\sim$}}}
\newcommand{\ltorder}{\mathrel{\raise.3ex\hbox{$<$}\mkern-14mu
            \lower0.6ex\hbox{$\sim$}}}

\shorttitle{A limit to relaxed systems without black holes}
\shortauthors{}

\begin{document}

\title{An upper limit to the velocity dispersion of relaxed
stellar systems without massive black holes}

\author{M. Coleman Miller\altaffilmark{1} and Melvyn B. Davies\altaffilmark{2}}

\altaffiltext{1}{Department of Astronomy and Joint Space-Science Institute,
University of Maryland, College Park, MD 20742-2421, USA}
\altaffiltext{2}{Lund Observatory, Department of Astronomy
and Theoretical Physics, Box 43, SE--221 00, Lund, Sweden}

\begin{abstract}

Massive black holes have been discovered in all closely examined
galaxies with high velocity dispersion.  The case is not as clear for
lower-dispersion systems such as low-mass galaxies and globular
clusters.  Here we suggest that above a critical velocity dispersion
$\sim 40$~km~s$^{-1}$, massive central black holes will form in
relaxed stellar systems at any cosmic epoch.  This is because above
this dispersion primordial binaries cannot support the system against
deep core collapse.  If, as previous simulations show, the black
holes formed in the cluster settle to produce a dense subcluster,
then given the extremely high densities reached during core collapse
the holes will merge with each other.  For low velocity dispersions
and hence low cluster escape speeds, mergers will typically kick out
all or all but one of the holes due to three-body kicks or  the
asymmetric emission of gravitational radiation.  If one hole remains,
it will tidally disrupt stars at a high rate.   If none remain, one
is formed after runaway collisions between stars, then it tidally
disrupts stars at a high rate.  The accretion rate after disruption
is many orders of magnitude above Eddington.  If, as several studies
suggest, the hole can accept matter at that rate because the
generated radiation is trapped and advected, then it will grow
quickly and form a massive central black hole.

\end{abstract}


\section{Introduction}

Observations over the last two decades have revealed central
massive black holes in all sufficiently well-observed massive
galaxies (e.g., \citealt{2011ApJ...738...17G}).  
However, the case is not as clear for lower-mass
galaxies or globular clusters, and indeed although there is 
evidence for black holes in some low-mass galaxies 
\citep{2010ApJ...721...26G,2011ApJ...727...20K} there
are examples of galaxies that clearly do not have black holes
that follow the standard mass -- velocity dispersion ($M-\sigma$)
relation \citep{2001Sci...293.1116M,2001AJ....122.2469G} 
and the case for globular clusters is far from
clear (e.g., \citealt{2002AJ....124.3270G, 2003ApJ...595..187M, 
2003ApJ...582L..21B,2012ApJ...750L..27S}). 

Here we approach this question by focusing on the velocity
dispersion rather than the mass of a stellar system.  In Section 2
we show that above a critical velocity dispersion $\sigma_{\rm
crit}\sim 40~{\rm km~s}^{-1}$, the total binding energy in
primordial binaries that can be tapped in three- and four-body
interactions is significantly less than the total binding energy of
the system as a whole, and hence if such systems are dynamically
relaxed they will undergo deep core collapse essentially unhindered
by dynamical heating from binaries (thus leading to one of the
scenarios discussed by \citealt{1978MNRAS.185..847B} in the context
of more massive clusters). We note that the galaxies seen thus far
without massive black holes have velocity dispersions below this
limit (e.g., NGC 205 has $\sigma=39$~km~s$^{-1}$ and M33 has
$\sigma=24$~km~s$^{-1}$; see \citealt{2009ApJ...698..198G} and
references therein). In Section 3 we discuss the evolution of
binary-free systems.  Previous studies have demonstrated that the
black holes in such systems sink rapidly to the center and interact
mostly with each other in a dense subcluster.   This leads to three
paths, all of which culminate in the formation of a massive black
hole: (1)~For  sufficiently high escape speed systems dynamical
interactions result in runaway merging of the black holes into a
massive hole.  For lower escape speed systems either one or zero
black holes remain after ejection of merged pairs due to asymmetric
emission of gravitational radiation during coalescence or Newtonian
recoil from interactions of black holes with dynamically formed
binaries.  (2)~If one black hole remains then it tidally disrupts
ordinary stars and consumes the remnant disks quickly, hence grows
rapidly into a massive black hole; other growth mechanisms, such as
the accretion of nascent gas or winds, are insignificant.  (3)~If no
black holes remain then runaway collisions form a massive star that
evolves into a black hole, and this first black hole grows via
accumulation of tidally disrupted stars.  Thus once binary support
is removed, massive black hole formation is assured as long as holes
consume tidal remnants quickly.  In Section 4 we determine the
minimum mass of a black hole formed via these paths and discuss the
implications of this scenario.

\section{Velocity dispersion threshold for deep core collapse}

Stellar systems that are in virial equilibrium evolve
via two-body interactions over their relaxation time, which 
for a star of mass $m$ in a system of velocity dispersion
$\sigma$ at a location with an average stellar mass density
$\rho$ is
\begin{equation}
t_{\rm rlx}\approx {0.3\over{\ln\Lambda}}{\sigma^3\over{
G^2\rho m}}
\end{equation}
\citep{1987degc.book.....S}, 
where $\ln\Lambda\sim 5-10$ is the Coulomb logarithm.  The
evolution of an isolated stellar system is towards a greater
concentration of stars in the center balanced by a greater
expansion of the cluster on the outskirts; there is a productive
analogy with thermodynamics, in which this behavior can
be seen as the gradual increase of cluster entropy (the greater
phase space accessed by the outer stars more than makes up for
the diminished phase space accessed by the stars in the core). 
It was demonstrated several decades ago that if all the stars
are single (as opposed to being in binary or multiple systems),
then over a timescale that scales with the relaxation time at
the half-mass radius for a typical star (where the multiple is
$\sim 15$ for an initially Plummer sphere of equal-mass stars
but is $\sim 0.2$ if there is a broad initial mass function; see
\citealt{2002ApJ...576..899P}), the core becomes so dense that
it loses thermal contact with the rest of the cluster and the
core undergoes a collapse such that the number density in the
inner portions scales as $n\sim r^{-2.2}$
\citep{1980MNRAS.191..483L,1980ApJ...242..765C}.  If we take
present-day nuclear star clusters as an example, then from
Figure~1 of \citet{2009ApJ...694..959M} we find that most
have half-mass relaxation times less than ${\rm few}\times 10^{10}$~yr
and thus are candidates to collapse within a Hubble time if
they had broad initial mass functions and no central massive
object to supply energy.

Binaries are the key to sustaining a cluster against this collapse.
When number densities become high enough that binary-single
interactions are common, such interactions can harden the binary
and hence inject energy into the cluster that decreases its density.
Many calculations (see, e.g., \citealt{1961AnAp...24..369H,
1975MNRAS.173..729H} for pioneering work) have shown
that binaries that are initially hard (meaning that their binding
energy exceeds the kinetic energy of a typical single star) tend
to harden via binary-single interactions, whereas initially soft
binaries tend to soften and eventually break up.  Consistent with
this expectation, globular clusters have a significantly smaller
binary fraction than is seen in the field 
(e.g., \citealt{1997ApJ...474..701R,2012A&A...540A..16M}).

In principle, even a very small number of binaries could have 
enough binding energy to hold off the collapse of a cluster.
Consider for example a reasonably rich globular cluster with a
velocity dispersion of 10~km~s$^{-1}$.  A binary of two solar-mass
stars near contact, with an orbital radius of 0.01~AU, has
$\sim 10^3$ times the binding energy per mass that a single cluster star
has in kinetic energy, so if 0.1\% of stars are in such binaries
the energy to hold off cluster collapse appears to be present.
White dwarfs are 100 times smaller yet, so it might seem that if there
is one near-contact white dwarf binary in a cluster of $10^5$ stars
its binary interactions could successfully oppose core collapse.

This is of course not true, for two reasons.  First, as the
semimajor axis of a binary shrinks, its close interactions with
single stars have a greater and greater chance of destroying the
single star or one of the binary stars, hence the kinetic energy
of recoil is not shared with the cluster \citep{1994ApJ...424..870D}.  
As an example, a tight
white dwarf binary cannot eject a main sequence star in this way.
Second, even if a three-body interaction is clean, a star 
that is thrown completely from the cluster cannot
share its kinetic energy with the cluster and the only expansion
of the core comes from the comparatively minor effect that the
core now has lost one star's worth of mass.  

The available binding energy from binaries is thus limited; clusters
having higher velocity dispersions  having a more limited
available binding energy.  As we now argue, this means that above
a velocity dispersion $\sigma_{\rm crit}\sim 40$~km~s$^{-1}$, 
the binaries cannot hold off core collapse. 
It should be noted that the velocity dispersion of a cluster will evolve
as a function of time, with velocity dispersions being somewhat larger
in the past when the cluster was more massive 
(e.g., \citealt{2009MNRAS.395.1173G,2010MNRAS.407.2241K}).
The effect could be particularly enhanced for clusters containing 
multiple stellar populations where a large fraction of the first generation
of stars are lost \citep{2008MNRAS.391..825D}. However the velocity
dispersion at later times will be more relevant to the discussion in this paper,
as this is when core collapse may typically be possible (i.e., on timescales
longer than the half-mass relaxation time).

As a first estimate of the available binding energy for a binary
with initial semimajor axis $a_0$, we assume that the eccentricity
distribution of binaries with a given semimajor axis is a thermal
distribution $P(e<e_0)=e_0^2$ truncated at the maximum
eccentricity $e_{\rm max}$ allowed for pericenter distances
greater than some minimum $r_{\rm p,min}$ (this could be the
pericenter distance at which stars collide), $a(1-e_{\rm
max})=r_{\rm p,min}$ for a semimajor axis $a$. Thus a fraction
$e_{\rm max}^2$ of orbits are allowed, hence the binding energy
that can be released from semimajor axis $a+da$ to $a$ is weighted
by $e_{\rm max}^2(a)=1-2r_{\rm
p,min}/a+(r_{\rm p,min}/a)^2$.  Thus the total available binding
energy from an initial semimajor axis $a_0$ with stars of mass $m$ is
\begin{equation}
E_{\rm bind,tot}(a_0)=\int_{r_{\rm p,min}}^{a_0}
{Gm^2\over {2a^2}}e^2_{\rm max}(a)da\; .
\end{equation}
This gives
\begin{equation}
\begin{array}{rl}
E_{\rm bind,tot}(a_0)&={Gm^2\over{2r_{\rm p,min}}}
\biggl[{1\over 3}-{r_{\rm p,min}\over a_0}+\left(r_{\rm p,min}\over a_0\right)^2\\
&-{1\over 3}\left(r_{\rm p,min}\over a_0\right)^3 \biggr] \; .
\end{array}
\label{eq:bind}
\end{equation}
For $a_0>10r_{\rm p,min}$, $E_{\rm bind,tot}$ is roughly constant at 
$Gm^2/6r_{\rm p,min}$ whereas it decreases rapidly below
$10r_{\rm p,min}$, so for simplicity we will approximate $E_{\rm bind,tot}$
as zero below $10r_{\rm p,min}$ and $Gm^2/6r_{\rm p,min}$ above it.

Our next step is to note that for stars formed in a low-density
environment, there is roughly one binary per single star, and the
binary semimajor axes are approximately equally distributed in $\ln
a$ from 0.01~AU to $\sim 10^4$~AU \citep{1982Ap&SS..88...55P}.  
In an environment where binaries
beyond a certain semimajor axis are ionized by binary-single
encounters, the fraction of binaries will be decreased.  For
example, if we begin with six single stars and six binaries and
ionize the ones larger than 1~AU, we now have fourteen single
stars and two binaries.  If as above we now only concentrate on
the binaries larger than $0.1~{\rm AU}=10r_{\rm p,min}$, this
represents $f\sim 7\%$ of the stars in the system.  Thus the binary
binding energy per {\it all} stars in the system is 
\begin{equation}
e_{\rm bin}/{\rm star}=fGm^2/(6r_{\rm p,min})\; .
\end{equation}
This is to be compared with the binding energy per star in the
cluster, which by the virial theorem equals the kinetic energy per
star in the cluster, or 
\begin{equation}
e_{\rm cluster}/{\rm star}={1\over 2}m\sigma^2
\end{equation}
for a velocity dispersion $\sigma$.  The point at which
$e_{\rm bin}/{\rm star}<e_{\rm cluster}/{\rm star}$ is the point at
which core collapse is theoretically possible.  From the numbers
above, if the single stars and the binary components both have
masses $\approx 1 M_\odot$ this happens when $\sigma\sim 40$~km~s$^{-1}$,
meaning that interactions with binaries of semimajor axis 
$\gtorder 0.5$~AU have positive total energy and are thus soft,
core collapse can proceed.
If the initial distribution of binary binding energies is extremely
unusual, e.g., if most stars are formed in binaries with semimajor
axes less than 0.5~AU, then the supply of binary energy would be
greater and the threshold velocity dispersion could in principle
be raised.  Barring such an unexpected distribution, however, the
threshold should be robust.

Indeed, work by \citet{1996IAUS..174..263C} suggests that there may
be less binary energy available than we derive above.
They take into account that,
rather than simply resetting the eccentricity of a binary, a
binary-single encounter can be resonant and hence for a given
interaction there is a greater chance to get to a very small
separation.  From their Figure~4 we infer that for solar-type stars and
$\sigma=40$~km~s$^{-1}$ a typical energy
$\Delta E\approx 6\times 10^{46}$~erg can be extracted from an
initially hard binary, whereas equation (\ref{eq:bind}) gives roughly an
order of magnitude larger energy.  Thus at $\sigma=40$~km~s$^{-1}$,
and perhaps at a slightly lower velocity dispersion, the energy that can
be extracted from primordial binaries is significantly less than
the binding energy of the cluster, hence such clusters can undergo
core collapse without being impeded significantly (three-body binary
formation and two-body tidal capture are also insignificant; see
\citealt{1983ApJ...268..319H} and \citealt{1975MNRAS.172P..15F}, respectively).

\section{Paths towards massive black hole formation}

\begin{figure}[htb]
\begin{center}
\hspace*{-1.3cm}
\includegraphics[scale=0.45]{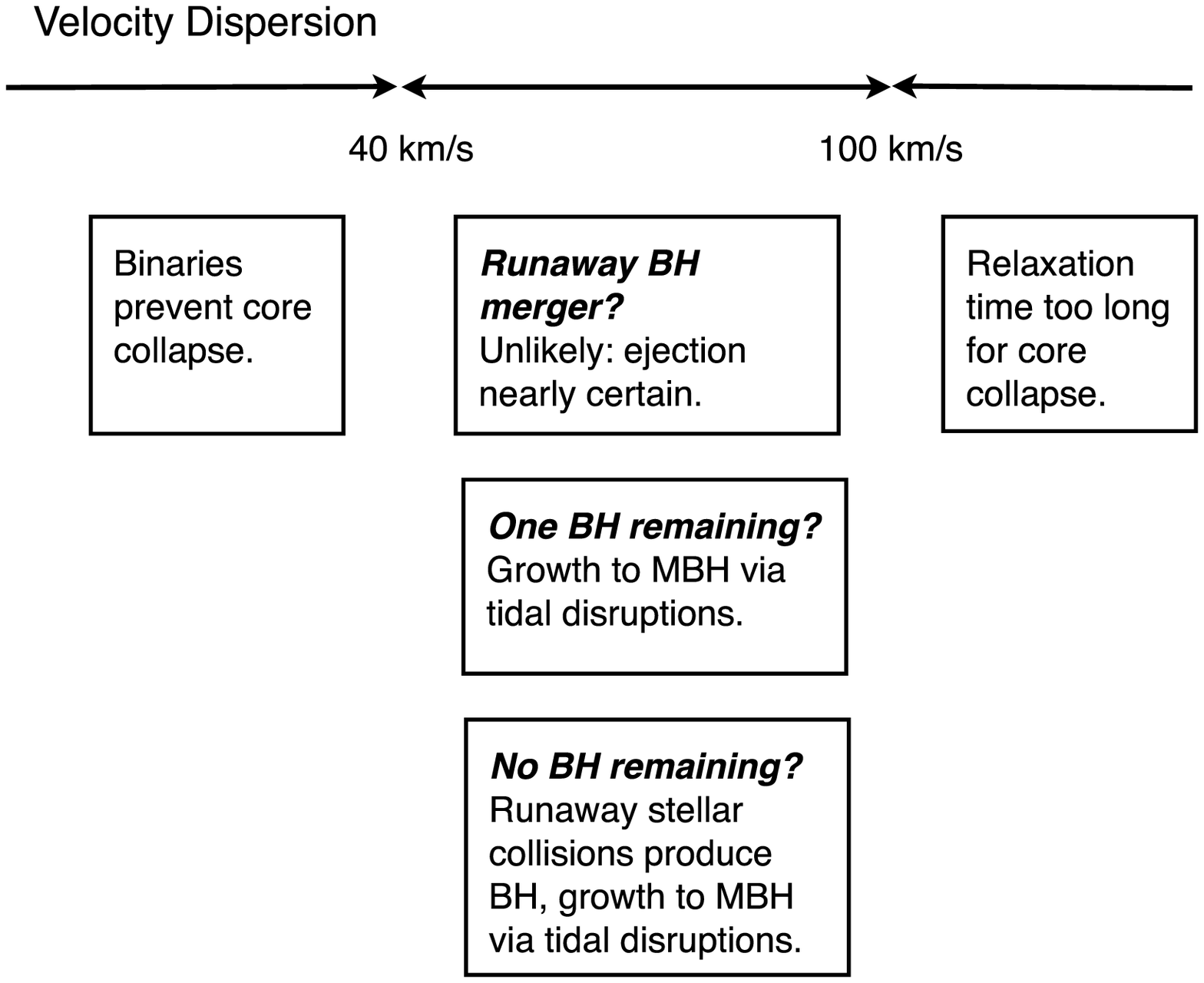}
\caption{Paths towards massive black hole formation in a stellar
cluster.  At velocity dispersions less than $\sim 40~{\rm km~s}^{-1}$,
heating from binaries prevents full core collapse.  At velocity
dispersions greater than $\sim 100~{\rm km~s}^{-1}$, a typical
nuclear star cluster will have too long a relaxation time for
its core to collapse in a Hubble time, although a massive black
hole can form in other ways.  At intermediate velocity
dispersions, full core collapse will occur and will likely result
in either zero or one remaining stellar-mass black hole.  In the
latter case, the hole will grow via tidal disruption of stars;
in the former, the stars will undergo runaway collisions that
produce a black hole, which will then grow via tidal disruption
of stars.}
\label{fig:flowchart}
\end{center}
\end{figure}

We now evaluate the paths towards massive black hole formation that we
mentioned in the introduction: runaway merging of black holes, tidal
disruption of stars by a single remaining black hole, and formation of
a new black hole from runaway collisions of stars, followed by tidal
disruption of stars by that black hole.   The first important question
is one of time scales.  In all three paths, the overwhelmingly longest
phase is the initial progression to core collapse.  To see this, note
that the time to core collapse is a multiple less than unity ($\sim
0.2$ for systems with a broad mass distribution; see
\citealt{2002ApJ...576..899P}) of the relaxation time of the nuclear
star cluster, which from Figure~1(a) of \citet{2009ApJ...694..959M} is
$t_{\rm rlx}\sim 10^9~{\rm yr} (M/10^6~M_\odot)$ with a spread of a
factor $\sim 10$.  There is some evidence that nuclear star clusters
obey a similar $M-\sigma$ relation to that seen for higher-mass black
holes.  There is some  observational evidence for this; e.g., Figure~2
from \citealt{2006ApJ...644L..21F}) indicates that nuclear star
clusters might have the same $M-\sigma$ slope as has been found for
black holes (see \citealt{2009ApJ...698..198G} for a recent discussion
of this relation) but offset so that the mass is a factor of $\sim 10$
higher than the black hole mass would be.  If we loosely equate the
velocity dispersion of the cluster with  that of the surrounding
bulge, this gives a cluster mass of
\begin{equation}
M_{\rm cl}\approx 10^6~M_\odot(\sigma/40~{\rm km~s}^{-1})^4
\end{equation}
based on the scalings of \citet{2009ApJ...698..198G}.  Thus
clusters with $\sigma\ltorder 100~{\rm km~s}^{-1}$ have a chance
to undergo core collapse within a Hubble time.  Figure~\ref{fig:flowchart}
illustrates the paths we consider.
 
This step is necessary for all three paths we discuss. For a
collapsed core, the self-similarity arguments of
\citet{1980MNRAS.191..483L}, and the classic simulations of
\citet{1980ApJ...242..765C}, show that if all the stars are treated
as point masses and no three-body binary formation is allowed, then
the density of a single-mass system evolves towards a $n\propto
r^{-2.2}$ configuration.  This is quite close to a singular
isothermal sphere $n\propto r^{-2}$, hence we will assume that the
velocity dispersion is nearly constant in the collapsed region.  As
a result, the relaxation time scales roughly as $\rho^{-1}$, so the
evolution timescale is shorter by orders of magnitude in the core of
the cluster  than it is in the cluster as a whole. 

As a result, once the core collapses all three paths are traveled in
a time much shorter than the time to collapse.  For example, runaway
mergers between black holes (or, in the third path, runaway
collisions between stars) occur roughly on the core relaxation
timescale, because when the number density is not yet sufficient for
frequent mergers or collisions, further relaxation will increase the
density on the relaxation time until interactions are frequent. Thus
the only limiting factor is the initial collapse time. We also note
that unlike in the scenario of runaway collapse of young massive
clusters proposed by \citet{2002ApJ...576..899P}, the time window
for runaway collisions of stars to form a single black hole (in the
third path) is not millions of years, but billions of years.  The
reason is that when the cluster is young enough that initially all
stars are on the main sequence, supernovae from the most massive
stars begin at $\sim$2.5~Myr and proceed for many stars, causing the
core to lose a large amount of mass to the ejecta and therefore
expand and lower the number density.  In contrast, in our picture
the evolution to core collapse is much later, perhaps billions of
years, hence the remaining stars are low-mass and thus only the
collision product will be massive enough to explode; very little
mass is lost, so the density remains high.

In addition to the general core collapse, in a multimass
system there is considerable mass segregation.  This means 
that the stars in the core will tend to be towards the
massive end, perhaps $\sim 1~M_\odot$ after billions of years.
In addition, of the objects
likely to be present after a long time, stellar-mass black
holes will be by a factor of a few to several the most massive.
Many studies (e.g., \citealt{2008MNRAS.386...65M}) have 
concluded that the black holes then form
a dense subcluster in which the holes interact mainly with themselves.
If, as in our scenario, there are no binaries, then the holes
can reach extremely high density in the center of the subcluster
and capture each other via emission of gravitational radiation
in initially hyperbolic two-body encounters.  
From \citet{1989ApJ...343..725Q}, the critical pericenter for
a two-body gravitational wave capture between two black
holes with a total mass $M=m_1+m_2$  and a reduced 
mass $\mu=m_1m_2/M$ is 
\begin{equation}
\begin{array}{rl}
r_{\rm p,GW}&=8.5\times 10^8~{\rm cm}~(M/20~M_\odot)^{5/7}\\
&\times(\mu/5~M_\odot)^{2/7}(\sigma/40~{\rm km~s}^{-1})^{-4/7}\; ,
\end{array}
\end{equation}
and hence their gravitationally focused cross section is
\begin{equation}
\begin{array}{rl}
\Sigma_{\rm bh}&=2\pi r_pGM/\sigma^2\approx
9\times 10^{23}~{\rm cm}^2(M/20~M_\odot)^{12/7}\\
&\times(\mu/5~M_\odot)^{2/7}
(\sigma/40~{\rm km~s}^{-1})^{-18/7}\; .
\end{array}
\end{equation}
When two black
holes capture each other in this way, their inspiral is extremely
rapid: from \citet{1964PhRv..136.1224P}, the inspiral time is
\begin{equation}
\begin{array}{rl}
a/(da/dt)&={5\over{64}}{c^5a^4(1-e^2)^{7/2}\over{
G^3\mu M^2(1+73e^2/24+37e^4/96)}}\\
&\approx 10^5~{\rm yr}
(a/R_\odot)^4(1-e^2)^{7/2}
\end{array}
\end{equation}
where the approximation is for $e\approx 1$ and our given number 
assumes $m_1=m_2=10~M_\odot$.  For $a\approx 1
R_\odot=7\times 10^{10}$~cm and $e>0.99$ (so that $r_p<r_{\rm p,GW}$),
the inspiral time is therefore less than 0.1~years, and for a
fixed $r_p$ the inspiral time scales as $a^{1/2}$ so that even
for an initial $a=100$~AU the inspiral time is just a few years.

When the holes do merge they emit gravitational radiation that is
in general asymmetric, meaning that the remnant single black hole
will recoil relative to its original center of mass.  Studies of
black hole recoil \citep{2007ApJ...668.1140B,2008ApJ...682L..29B,
2008PhRvD..77d4028L,2009PhRvD..79f4018L,2010CQGra..27k4006L,
2010ApJ...719.1427V,2012arXiv1201.1923L} show that although kicks from 
the coalescence of nonspinning black holes are limited to $<200~{\rm
km~s}^{-1}$, rapidly spinning black holes can produce remnants
that travel at thousands of kilometers per second relative to
their original center of mass. Thus in this environment, unlike in
the conditions that may exist in the $z>10$ universe 
\citep{2011ApJ...740L..42D},
mergers that are restricted to comparable-mass black holes are
most likely to lead to an ejection of the remnant.  

As pointed out
to us by S. Sigurdsson (private communication), for low velocity
dispersions the ejection of black holes is likely to be dominated
by encounters with hard binaries formed by the interaction of three
initially hyperbolic black holes.  \citet{1985IAUS..113..231H}
finds that the rate of formation of ``immortal" binaries by
this process (i.e., binaries that are not later softened and
ionized) is ${\dot n}_{3B}=126G^5m^5n^3/\sigma^9$, where he
assumes objects of identical mass $m$ and $\sigma$ is the three-dimensional
velocity dispersion.  Thus the ratio of the formation rate per
volume of these binaries to the rate of gravitational wave capture
of black holes by each other, assuming equal masses, is
\begin{equation}
{{\dot n}_{3B}\over{{\dot n}_{\rm BHBH}}}\approx
200 (n/10^{10}~{\rm pc}^{-3})(m/10~M_\odot)^3
(\sigma/40~{\rm km~s}^{-1})^{-52/9}\; .
\end{equation}
Thus for low to moderate velocity dispersions, and high number
densities, binary-single ejections are likely to dominate.  The
result will be similar to the case in which only double black
hole mergers occur: there will either be zero or one hole left.
For the rest of this section we concentrate on ejections by mergers.

We set up a simple simulation of the evolution of a black hole
subcluster with no binaries. We assume that initially there are
either 100 or 101 black holes; note that even if all mergers eject
the remnant, having an odd number initially guarantees that one
will survive because with no binaries the interactions are
pairwise.  The distribution of black hole masses is not
well-established, and the distribution of their spins is even less
so, but as an illustrative example we draw the initial masses of
the black holes from the range $[5,30)~M_\odot$, with a
distribution $dN/dM\propto M^{-2}$, and the initial spins are drawn
uniformly from the range $cJ/(GM^2)=[0,1)$.

We simulate the evolution of the cluster interaction by
interaction using the rejection method: we select two black
holes randomly, compute the cross section $\Sigma$ of the interaction,
divide by the largest possible cross section $\Sigma_{\rm max}$
(which is the cross section of capture by the two most
massive black holes in the sample), and then compare that
ratio with a uniform random deviate $x\in [0,1)$.  If 
$x<\Sigma/\Sigma_{\rm max}$ we accept the interaction,
otherwise we draw again.

If the interaction is between two black holes, then we use
the recent \citet{2012arXiv1201.1923L} formula for the kick.  If
the kick is greater than the escape speed $v_{\rm esc}=4\sigma$ 
(typical of a core-collapsed cluster) we assume that the remnant
has been ejected from the cluster and thus we remove both
black holes from the sample.  Otherwise, we assume
the remnant remains, hence we sum the masses of the holes
and estimate the spin of the remnant following the prescription
given in \citet{2008PhRvD..78d4002R}.  

\begin{figure}[htb]
\begin{center}
\plotone{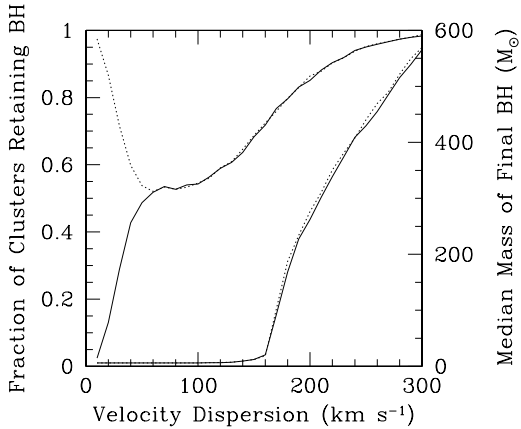}
\caption{Fraction of clusters of a given velocity dispersion
that retain a black hole after a succession of mergers (upper
left curves) and, if a black hole is left, the median mass  of
the remaining black hole (lower right curves).  For this figure
we ignore the effects of hard binaries formed by the
interactions of three initially hyperbolic black holes (see
text), hence there is a difference between cases with an
initially even and an initially odd number of black holes. The
solid curves are for 100 initial black holes, and the dotted
curves are for 101 initial black holes; the asymmetry in
retained fraction at low velocity dispersions is because if
every black hole merger results in an ejection, an initially
even number will leave behind no black holes whereas an
initially odd number will leave behind one.  We assume an escape
speed that is four times the velocity dispersion.  This figure
demonstrates that retained runaway mergers leading to massive
seeds are only likely for velocity dispersions $\gtorder
200$~km~s$^{-1}$.}
\label{fig:bh}
\end{center}
\end{figure}

Figure~\ref{fig:bh} shows the results.  Here we
plot the fraction of clusters that retain a black hole
after subcluster evolution, and the median mass of the
final black hole if one remains, as a function of the velocity
dispersion $\sigma$ of the cluster.  For each velocity dispersion
we performed $10^4$ simulations.  For low escape speeds, almost
all mergers between black holes eject the remnant, hence
retention depends on whether the initial number of holes is
even or odd.  As the escape speed increases, so does the
probability that a merger will not eject the remnant; for
$v_{\rm esc}\ltorder 100$~km~s$^{-1}$ it is most probable
that this happens when the spins of the holes are low and
their masses are close to each other (note from symmetry 
that there is zero recoil from the merger of equal-mass
nonspinning holes).  As the escape speed increases further,
mergers between black holes of different masses can be
retained, until at $v_{\rm esc}\gtorder 800$~km~s$^{-1}$
a runaway occurs and a single victorious black hole is 
usually the result.

From these simulations we can argue that for clusters with
velocity dispersions $\ltorder 100$~km~s$^{-1}$ a runaway
is unlikely, but that there is roughly an equal chance of
leaving behind either one or zero holes (depending largely
on the parity of the initial number until $\sigma\gtorder
60$~km~s$^{-1}$).  When there is a black hole left behind
it is likely for $\sigma\ltorder 100$~km~s$^{-1}$ to be
at the low end  of the mass distribution ($\sim 5~M_\odot$),
because  such black holes are initially more common.  In addition, 
lower-mass black holes have a lower cross section for capture
and hence an enhanced probability of survival.

The subsequent evolution has two possibilities:

{\it One black hole remains.}---Then as we discuss in the next
section, the black hole will sit near the center of the high
number density distribution of stars.  Tidal disruptions will
add a few tens of percent of the stellar mass to the hole, mostly
within a few weeks or less of the initial disruption, hence the
hole will grow quickly.  Given that interactions with the stars
cannot eject the hole from the cluster, it will become a massive
black hole in a short timescale.

{\it No black holes remain.}---In this case, the stars will
undergo runaway collisions with themselves, leading to the
production of a massive star that will then become a black hole
(e.g., \citealt{2002ApJ...576..899P}).  The situation then
reduces to the previous case, because the time needed to produce
a {\it second} black hole, which could potentially eject the
first, is significantly larger than the time needed for the
first hole to increase its mass to the point that it can no
longer be ejected.

We now discuss these possibilities in greater depth.

\subsection{Interactions between stars and a black hole}

Although the central density after core collapse is formally
infinite, the finite number of stars means that this translates
to a few stars in a small region near the core.  For example,
if we consider the inner $\sim 10$ solar-type stars after core
collapse and continue to assume a constant velocity dispersion,
then they are in a region $r=GM/\sigma^2\sim 5~{\rm AU}
(M/10~M_\odot)(\sigma/40~{\rm km~s}^{-1})^{-2}$ in radius,
with a resulting number density of $n>10^{14}~{\rm pc}^{-3}$.
Even the inner 1000 stars are in a region with $n>10^{10}~{\rm pc}^{-3}$,
so interactions will be common and rapid.

{\it Stellar tidal disruption by black holes.}---A
promising mechanism for such runaway growth is tidal disruption
of stars by stellar-mass black holes.  
The critical pericenter for tidal disruption of a star of
mass $m$ and radius $R$ by a black hole of mass $M$ is
\begin{equation}
r_{\rm p,tidal}=(3M/m)^{1/3}R\; .
\end{equation}
Thus the gravitationally focused cross section for tidal disruption,
assuming that the black hole mass greatly exceeds the stellar mass,
is
\begin{equation}
\Sigma_{\rm tidal}\approx 10^{26}~{\rm cm}^2(M/10~M_\odot)^{4/3}
(\sigma/40~{\rm km~s}^{-1})^{-2}
\end{equation}
for solar-type stars.  This is roughly an order of magnitude
greater than the star-star collision cross section discussed
later, and two orders
of magnitude larger than the black hole -- black hole capture
cross section.  Moreover, the rate is nonlinear in the mass of the
black hole ($\Sigma\propto M^{4/3}$).  Thus the conditions
for a runaway exist.

If tidal disruption does occur, then the mass will be force fed
to the black hole at an extremely super-Eddington rate.  
Studies suggest that fallback initially occurs over several
times the internal dynamical time of the disrupted star 
\citep{1989ApJ...346L..13E},
which is several hours for a solar-type star.  The accretion
rate is therefore many millions of times
the Eddington rate of a stellar-mass black hole.  Analyses of such
supercritical accretion (e.g., \citealt{1976ApJ...206..295M,
1979MNRAS.187..237B,1980AcA....30....1J,1999ApJ...518..356P,
2005ApJ...628..368O}) indicate that the matter will indeed
flow into the hole at that rate, but that most of the photon luminosity
that is generated will be advected in with the very optically thick
matter (hence although the accretion rate is tremendously 
super-Eddington, the luminosity could be limited to Eddington or
slightly higher).  Thus it is expected that within a matter of days, i.e.,
much shorter than any other relevant timescale, most of the bound
remainder of the star will flow onto the hole.  If this is the case,
then the majority of the accretion will finish without harassment
from additional encounters by stars.  If, on the contrary, the
accretion rate is actually limited to the Eddington rate then the
time needed to accrete most of the matter is much longer than the
time to the next encounter, and the disk might be disrupted, leading
to negligible growth of the hole.

The unbound remnant of the star will be thrown outwards at speeds
comparable to the orbital speed at tidal disruption, which is
$\sim 800~{\rm km/s}(M/10~M_\odot)^{1/3}$ for a solar-type star.
This is much larger than the escape speed, so the wind will depart
ballistically unless it runs into many times its own mass in gas
in the cluster.  However, given that the virial temperature of the 
cluster is $\sim 10^5~{\rm K}(\sigma/40~{\rm km~s}^{-1})^2$ and that 
cooling is extremely efficient at that temperature, the total
amount of gas in the cluster at a given time will be small even
though its escape speed is sufficient to retain winds from red
giants or (earlier, when more massive stars existed) planetary
nebulae.  Thus we assume that the unbound gas simply escapes from
the cluster.  The ratio of unbound gas to gas that accretes onto
the black hole is rather uncertain.  The initial disruption leaves
about half the mass bound \citep{1989ApJ...346L..13E}, but shocks upon
the return of the bound matter might unbind additional mass.  In a
recent study by \citet{2009MNRAS.400.2070S} they consider different
ejection fractions
ranging from $f_{\rm esc}=0.5$ (corresponding to negligible return shocks)
to $f_{\rm esc}=0.8$ (corresponding to powerful return shocks).
In our scenario, the upshot is that because a single black hole
will grow, its growth will eject up to a few times its own mass
in stellar debris.  Until this reaches at least hundreds, and probably
thousands, of solar masses this will be such a small fraction of
even the core mass that we expect it to have a minor effect on the
dynamical evolution.

{\it Star-star collisions.}---At the velocity dispersions we consider,
these collisions are likely to lead to mergers with little mass loss, 
because $\sigma\sim 40~{\rm km~s}^{-1}$ is much less than the escape
speed $\sim 600~{\rm km~s}^{-1}$ of a solar-type star.  For
the same reason, these collisions are gravitationally focused,
with a cross section $\Sigma=\pi r_p^2(1+2GM_{\rm tot}/(r_p\sigma^2))\approx
2\pi(GM_{\rm tot}/\sigma^2)r_p$ for a pericenter distance $r_p$ and
a total mass between the stars of $M_{\rm tot}$.  The
relevant pericenter distance is the sum of the stellar radii,
which is $2R_\odot\approx 0.01$~AU for two solar-type stars, hence
for two such stars
$\Sigma\approx 1.5\times 10^{25}~{\rm cm}^2(\sigma/40~{\rm km~s}^{-1})^{-2}$.
The characteristic time of interaction is then
$\tau=1/(n\Sigma\sigma)\approx 10^6~{\rm yr}(n/10^{10}~{\rm pc}^{-3})^{-1}
(M_{\rm tot}/2~M_\odot)^{-1}(\sigma/40~{\rm km~s}^{-1})^{-1}$.
Note that as a result even for the inner $\sim 10^3$ stars the
collision time for solar-type stars is much less than their
$\sim 3\times 10^7$~yr Kelvin-Helmholtz time, hence the stars
will not be able to radiate their collisional energy before the
next collision.  However, because the velocity dispersion is
$<0.1$ times the stellar escape speed, the energy added is minor
and most of the pressure holding up the collision product stems
from gravitational contraction rather than either collision
energy or nuclear energy; these are thus not stars in the standard
sense, and need not have luminosities as high as those of main
sequence stars of the same mass.

In addition, on the main sequence, stellar radii increase with
increasing mass, hence the rate of interactions increases more than
linearly with increasing stellar mass.  An additional factor
is that more massive stars tend to sit closer to the center of
the potential, where the number density of objects is greater.
The conditions are thus ripe for a runaway, and indeed runaway
merging of stars has been proposed as a mechanism for the
generation of supermassive stars that later evolve into
intermediate-mass black holes \citep{2002ApJ...576..899P,
2006MNRAS.368..141F}.  It has been
suggested that the high wind rates expected for high-mass stars
can severely limit the growth of supermassive stars 
\citep{2009A&A...497..255G}.  Note,
however, that these wind rates are based on extrapolations of
winds for main sequence stars, and as indicated above the
collision products will be substantially larger and less
luminous than main sequence stars.  Indeed, the collision products
are more likely to be a ``bag of cores" than an actual star, where an 
extended gaseous envelope engulfs an ensemble of stellar cores. 

We do note that although \citet{2009A&A...497..255G} argue that
winds may prevent the formation of intermediate-mass black holes,
they find that runaway collisions produce stars massive enough
to evolve to normal stellar-mass black holes, at least.  Thus
for our purposes we assume that star-star collisions will lead
to black hole production.

The question is then whether the first black hole that forms
has enough time to consume many stars so that by the time the
next black hole forms, the first one is so massive that any
BH-BH merger will produce a weak recoil that retains the remnant
in the cluster.  We argue that this is in fact the case: the
first black hole to form will be at the center of the mass
distribution, where the number density is the highest.  If this
is, for example, in the region occupied by the inner $\sim 100$
stars, then the number density is such that tidal disruptions of
stars by even a $10~M_\odot$ black hole occur on average once
per few hundred years, and the interaction time scales as $M^{-4/3}$.  
Thus the hole will double its mass every few thousand years, i.e.,
in a time vastly shorter than the lifetime of even the most
massive stars.  We note that although the segregation of the 
black holes to the center of the cluster and their ejection 
leads to some flattening of the stellar number density near the
center of the cluster (see, e.g., \citealt{2012MNRAS.tmp.2546A}
and in particular his Figure~8 for a recent N-body simulation),
we expect that when most of the black holes have been ejected
the stars near the core, which have a very short relaxation time,
will regrow the cusp.
Hence the first black hole formed due to runaway
stellar collisions will be able to increase its mass by a large
factor before any other new-generation black hole forms.

\subsection{Minimum mass of central BH}

We consider here the evolution of a cluster with a velocity
dispersion large enough to guarantee core collapse.  If a
black hole grows in the cluster, what is a rough approximation
to its minimum mass? 
We will approach this question in two different ways.  First, we will
determine the mass of a black hole nailed to the center of an
$n\propto r^{-2}$ core collapse cluster such that dynamical
processes around  the black hole can supply enough heat to
help forestall further core collapse.  Second, we will apply
the criterion that the wander radius of the black hole must
be less than its radius of influence, under the assumption
that otherwise the number of stars bound to the black hole
would be much less, hence its heating influence would be
reduced.

We will assume as before that the mass of the nuclear star cluster
is related to its velocity dispersion by
$M_{\rm cl}\approx 10^6~M_\odot(\sigma/40~{\rm km~s}^{-1})^4$.  For a core
collapse cluster with $n\propto r^{-2}$, the velocity dispersion
is the same at all radii and the gravitational binding energy is
$E_{\rm bind}=(1/2)M_{\rm cl}\sigma^2$ from the virial theorem.

This energy must be compared with the available energy (as defined
before) from dynamics around the central black hole.  The available
energy per unit mass around the black hole that we found previously is
$GM_{\rm BH}/(6r_{\rm p,min})$, where $r_{\rm p,min}$ is the minimum
pericenter distance of an orbit that can last long enough for
significant dynamical interactions.  For a black hole, the relevant
time is the time for gravitational radiation to cause the object to
spiral in; this time scales as $T\sim (mM_{\rm BH}^2)^{-1}r_p^4$, roughly, so
for a fixed $T$ we have $r_{\rm p,min}\propto M_{\rm BH}^{1/2}$.  We
used $r_{\rm p,min}\approx 0.01$~AU for $10~M_\odot$ (giving an
inspiral time of a few million years), so we will adopt
$r_{\rm p,min}=0.1~{\rm AU}(M_{\rm BH}/10^3~M_\odot)^{1/2}$.

If the distribution of stars around the black hole is a steep cusp, then
the stellar mass in the radius of influence of the black hole equals
the mass of the black hole (this need not be true if the density
distribution has a core profile; see equation (14) of 
\citealt{2011ApJ...735...89L}).  When we compare the available dynamical
energy of the stars around the black hole with the binding energy
of the cluster, we find that
\begin{equation}
M_{\rm BH}\gtorder 500~M_\odot(\sigma/40~{\rm km~s}^{-1})^4
\end{equation}
is required for the stars around the black hole to provide sufficient
energy to hold off collapse.

We can also approach this from a different angle.  A finite-mass black
hole will not be nailed to the center of the cluster.  Instead, it
will wander due to stochastic dynamical interactions.  If the wander
radius is less than its radius of influence then we can suppose that
it is near the center of the stellar distribution where encounters
are frequent, but if the wander radius is larger then this need
not be the case and heating could be less efficient.  Thus a different
criterion is $r_{\rm wander}<r_{\rm infl,BH}$.  Suppose that there
is a nearly constant-density core in the inner 10\% of the
cluster; then the scale height of
a species in a cluster is inversely proportional to the square root
of its mass (from energy equipartition arguments), hence for this system
we expect
\begin{equation}
r_{\rm wander}\sim r_{\rm cl}(\langle m\rangle/M_{\rm BH})^{1/2}\; .
\end{equation}
Here $\langle m\rangle$ is the average mass of a star.  The cluster
radius is $r_{\rm cl}=GM_{\rm cl}/\sigma^2$ and the radius of influence
of the black hole is $GM_{\rm BH}/\sigma^2$, hence the wander
criterion is
\begin{equation}
\begin{array}{rl}
0.1(GM_{\rm cl}/\sigma^2)(\langle m\rangle/M_{\rm BH})^{1/2}&\ltorder
GM_{\rm BH}/\sigma^2\\
M_{\rm BH}&\gtorder(0.01~M_{\rm cl}^2\langle m\rangle)^{1/3}\\
M_{\rm BH}&\gtorder 2\times 10^3~M_\odot (\sigma/40~{\rm km~s}^{-1})^{8/3}\; ,\\
\end{array}
\end{equation}
where in the last line we assume $\langle m\rangle=1~M_\odot$.

Recall that these are {\it lower limits} on the mass of the central
black hole.  The mass could be considerably greater depending on 
long-term accretion of stars or gas.

\section{Discussion and conclusions}

We have discussed the evolution of a relaxed cluster that has
a velocity dispersion $\sigma\gtorder 40$~km~s$^{-1}$, which 
is large enough to render binaries insignificant,
but that does not initially contain a massive central black hole.
We argue that a massive hole will inevitably form if it can
swallow tidal debris rapidly: interactions
in the black hole subcluster will leave either zero or one
hole. In the case of zero, a black hole will form from the
product of runaway stellar merging.  In either case, the
hole will feed quickly from the remnants of the stars it tidally
disrupts, and hence will grow until it has significant dynamical
effects on the cluster and thus slows its own growth.  It is
not guaranteed that the holes will then follow the same
$M-\sigma$ relation that exists for higher velocity dispersion
systems.  It is also not guaranteed that clusters with lower
velocity dispersions will {\it not} have black holes, but it
is possible that massive black-hole formation
is prevented as long as binaries have a significant heating effect
(see \citealt{2008ApJ...686..303G} for a numerical exploration
of the heating due to binaries or a massive central object).

\acknowledgements

We thank Sverre Aarseth, Tal Alexander, Jillian Bellovary,
Kayhan G\"ultekin, David Merritt, and Steinn Sigurdsson for 
valuable discussions.  We also thank the referee for a
constructive report.
This work was supported by NASA ATP grants NNX08AH29G and
NNX12AG29G (MCM) and Swedish Research Council grants
2008--4089 and 2011--3991 (MBD).

\bibliography{bh}

\begin{thebibliography}{54}
\expandafter\ifx\csname natexlab\endcsname\relax\def\natexlab#1{#1}\fi

\bibitem[{{Aarseth}(2012)}]{2012MNRAS.tmp.2546A}
{Aarseth}, S.~J. 2012, \mnras, 2546

\bibitem[{{Baker} {et~al.}(2007){Baker}, {Boggs}, {Centrella}, {Kelly},
  {McWilliams}, {Miller}, \& {van Meter}}]{2007ApJ...668.1140B}
{Baker}, J.~G., {Boggs}, W.~D., {Centrella}, J., {Kelly}, B.~J., {McWilliams},
  S.~T., {Miller}, M.~C., \& {van Meter}, J.~R. 2007, \apj, 668, 1140

\bibitem[{{Baker} {et~al.}(2008){Baker}, {Boggs}, {Centrella}, {Kelly},
  {McWilliams}, {Miller}, \& {van Meter}}]{2008ApJ...682L..29B}
---. 2008, \apjl, 682, L29

\bibitem[{{Baumgardt} {et~al.}(2003){Baumgardt}, {Hut}, {Makino}, {McMillan},
  \& {Portegies Zwart}}]{2003ApJ...582L..21B}
{Baumgardt}, H., {Hut}, P., {Makino}, J., {McMillan}, S., \& {Portegies Zwart},
  S. 2003, \apjl, 582, L21

\bibitem[{{Begelman}(1979)}]{1979MNRAS.187..237B}
{Begelman}, M.~C. 1979, \mnras, 187, 237

\bibitem[{{Begelman} \& {Rees}(1978)}]{1978MNRAS.185..847B}
{Begelman}, M.~C. \& {Rees}, M.~J. 1978, \mnras, 185, 847

\bibitem[{{Chernoff} \& {Huang}(1996)}]{1996IAUS..174..263C}
{Chernoff}, D.~F. \& {Huang}, X. 1996, in IAU Symposium, Vol. 174, Dynamical
  Evolution of Star Clusters: Confrontation of Theory and Observations, ed.
  {P.~Hut \& J.~Makino}, 263

\bibitem[{{Cohn}(1980)}]{1980ApJ...242..765C}
{Cohn}, H. 1980, \apj, 242, 765

\bibitem[{{Davies} {et~al.}(1994){Davies}, {Benz}, \&
  {Hills}}]{1994ApJ...424..870D}
{Davies}, M.~B., {Benz}, W., \& {Hills}, J.~G. 1994, \apj, 424, 870

\bibitem[{{Davies} {et~al.}(2011){Davies}, {Miller}, \&
  {Bellovary}}]{2011ApJ...740L..42D}
{Davies}, M.~B., {Miller}, M.~C., \& {Bellovary}, J.~M. 2011, \apjl, 740, L42

\bibitem[{{D'Ercole} {et~al.}(2008){D'Ercole}, {Vesperini}, {D'Antona},
  {McMillan}, \& {Recchi}}]{2008MNRAS.391..825D}
{D'Ercole}, A., {Vesperini}, E., {D'Antona}, F., {McMillan}, S.~L.~W., \&
  {Recchi}, S. 2008, \mnras, 391, 825

\bibitem[{{Evans} \& {Kochanek}(1989)}]{1989ApJ...346L..13E}
{Evans}, C.~R. \& {Kochanek}, C.~S. 1989, \apjl, 346, L13

\bibitem[{{Fabian} {et~al.}(1975){Fabian}, {Pringle}, \&
  {Rees}}]{1975MNRAS.172P..15F}
{Fabian}, A.~C., {Pringle}, J.~E., \& {Rees}, M.~J. 1975, \mnras, 172, 15P

\bibitem[{{Ferrarese} {et~al.}(2006){Ferrarese}, {C{\^o}t{\'e}}, {Dalla
  Bont{\`a}}, {Peng}, {Merritt}, {Jord{\'a}n}, {Blakeslee}, {Ha{\c s}egan},
  {Mei}, {Piatek}, {Tonry}, \& {West}}]{2006ApJ...644L..21F}
{Ferrarese}, L., {C{\^o}t{\'e}}, P., {Dalla Bont{\`a}}, E., {Peng}, E.~W.,
  {Merritt}, D., {Jord{\'a}n}, A., {Blakeslee}, J.~P., {Ha{\c s}egan}, M.,
  {Mei}, S., {Piatek}, S., {Tonry}, J.~L., \& {West}, M.~J. 2006, \apjl, 644,
  L21

\bibitem[{{Freitag} {et~al.}(2006){Freitag}, {G{\"u}rkan}, \&
  {Rasio}}]{2006MNRAS.368..141F}
{Freitag}, M., {G{\"u}rkan}, M.~A., \& {Rasio}, F.~A. 2006, \mnras, 368, 141

\bibitem[{{Gebhardt} {et~al.}(2001){Gebhardt}, {Lauer}, {Kormendy}, {Pinkney},
  {Bower}, {Green}, {Gull}, {Hutchings}, {Kaiser}, {Nelson}, {Richstone}, \&
  {Weistrop}}]{2001AJ....122.2469G}
{Gebhardt}, K., {Lauer}, T.~R., {Kormendy}, J., {Pinkney}, J., {Bower}, G.~A.,
  {Green}, R., {Gull}, T., {Hutchings}, J.~B., {Kaiser}, M.~E., {Nelson},
  C.~H., {Richstone}, D., \& {Weistrop}, D. 2001, \aj, 122, 2469

\bibitem[{{Gerssen} {et~al.}(2002){Gerssen}, {van der Marel}, {Gebhardt},
  {Guhathakurta}, {Peterson}, \& {Pryor}}]{2002AJ....124.3270G}
{Gerssen}, J., {van der Marel}, R.~P., {Gebhardt}, K., {Guhathakurta}, P.,
  {Peterson}, R.~C., \& {Pryor}, C. 2002, \aj, 124, 3270

\bibitem[{{Giersz} \& {Heggie}(2009)}]{2009MNRAS.395.1173G}
{Giersz}, M. \& {Heggie}, D.~C. 2009, \mnras, 395, 1173

\bibitem[{{Gill} {et~al.}(2008){Gill}, {Trenti}, {Miller}, {van der Marel},
  {Hamilton}, \& {Stiavelli}}]{2008ApJ...686..303G}
{Gill}, M., {Trenti}, M., {Miller}, M.~C., {van der Marel}, R., {Hamilton}, D.,
  \& {Stiavelli}, M. 2008, \apj, 686, 303

\bibitem[{{Glebbeek} {et~al.}(2009){Glebbeek}, {Gaburov}, {de Mink}, {Pols}, \&
  {Portegies Zwart}}]{2009A&A...497..255G}
{Glebbeek}, E., {Gaburov}, E., {de Mink}, S.~E., {Pols}, O.~R., \& {Portegies
  Zwart}, S.~F. 2009, \aap, 497, 255

\bibitem[{{Greene} {et~al.}(2010){Greene}, {Peng}, {Kim}, {Kuo}, {Braatz},
  {Violette Impellizzeri}, {Condon}, {Lo}, {Henkel}, \&
  {Reid}}]{2010ApJ...721...26G}
{Greene}, J.~E., {Peng}, C.~Y., {Kim}, M., {Kuo}, C.-Y., {Braatz}, J.~A.,
  {Violette Impellizzeri}, C.~M., {Condon}, J.~J., {Lo}, K.~Y., {Henkel}, C.,
  \& {Reid}, M.~J. 2010, \apj, 721, 26

\bibitem[{{G{\"u}ltekin} {et~al.}(2009){G{\"u}ltekin}, {Richstone}, {Gebhardt},
  {Lauer}, {Tremaine}, {Aller}, {Bender}, {Dressler}, {Faber}, {Filippenko},
  {Green}, {Ho}, {Kormendy}, {Magorrian}, {Pinkney}, \&
  {Siopis}}]{2009ApJ...698..198G}
{G{\"u}ltekin}, K., {Richstone}, D.~O., {Gebhardt}, K., {Lauer}, T.~R.,
  {Tremaine}, S., {Aller}, M.~C., {Bender}, R., {Dressler}, A., {Faber}, S.~M.,
  {Filippenko}, A.~V., {Green}, R., {Ho}, L.~C., {Kormendy}, J., {Magorrian},
  J., {Pinkney}, J., \& {Siopis}, C. 2009, \apj, 698, 198

\bibitem[{{G{\"u}ltekin} {et~al.}(2011){G{\"u}ltekin}, {Tremaine}, {Loeb}, \&
  {Richstone}}]{2011ApJ...738...17G}
{G{\"u}ltekin}, K., {Tremaine}, S., {Loeb}, A., \& {Richstone}, D.~O. 2011,
  \apj, 738, 17

\bibitem[{{Heggie}(1975)}]{1975MNRAS.173..729H}
{Heggie}, D.~C. 1975, \mnras, 173, 729

\bibitem[{{H{\'e}non}(1961)}]{1961AnAp...24..369H}
{H{\'e}non}, M. 1961, Annales d'Astrophysique, 24, 369

\bibitem[{{Hut}(1985)}]{1985IAUS..113..231H}
{Hut}, P. 1985, in IAU Symposium, Vol. 113, Dynamics of Star Clusters, ed.
  {J.~Goodman \& P.~Hut}, 231--247

\bibitem[{{Hut} \& {Bahcall}(1983)}]{1983ApJ...268..319H}
{Hut}, P. \& {Bahcall}, J.~N. 1983, \apj, 268, 319

\bibitem[{{Jaroszynski} {et~al.}(1980){Jaroszynski}, {Abramowicz}, \&
  {Paczynski}}]{1980AcA....30....1J}
{Jaroszynski}, M., {Abramowicz}, M.~A., \& {Paczynski}, B. 1980, \actaa, 30, 1

\bibitem[{{Kuo} {et~al.}(2011){Kuo}, {Braatz}, {Condon}, {Impellizzeri}, {Lo},
  {Zaw}, {Schenker}, {Henkel}, {Reid}, \& {Greene}}]{2011ApJ...727...20K}
{Kuo}, C.~Y., {Braatz}, J.~A., {Condon}, J.~J., {Impellizzeri}, C.~M.~V., {Lo},
  K.~Y., {Zaw}, I., {Schenker}, M., {Henkel}, C., {Reid}, M.~J., \& {Greene},
  J.~E. 2011, \apj, 727, 20

\bibitem[{{K{\"u}pper} {et~al.}(2010){K{\"u}pper}, {Kroupa}, {Baumgardt}, \&
  {Heggie}}]{2010MNRAS.407.2241K}
{K{\"u}pper}, A.~H.~W., {Kroupa}, P., {Baumgardt}, H., \& {Heggie}, D.~C. 2010,
  \mnras, 407, 2241

\bibitem[{{Lasota} {et~al.}(2011){Lasota}, {Alexander}, {Dubus}, {Barret},
  {Farrell}, {Gehrels}, {Godet}, \& {Webb}}]{2011ApJ...735...89L}
{Lasota}, J.-P., {Alexander}, T., {Dubus}, G., {Barret}, D., {Farrell}, S.~A.,
  {Gehrels}, N., {Godet}, O., \& {Webb}, N.~A. 2011, \apj, 735, 89

\bibitem[{{Lousto} {et~al.}(2010){Lousto}, {Campanelli}, {Zlochower}, \&
  {Nakano}}]{2010CQGra..27k4006L}
{Lousto}, C.~O., {Campanelli}, M., {Zlochower}, Y., \& {Nakano}, H. 2010,
  Classical and Quantum Gravity, 27, 114006

\bibitem[{{Lousto} \& {Zlochower}(2008)}]{2008PhRvD..77d4028L}
{Lousto}, C.~O. \& {Zlochower}, Y. 2008, \prd, 77, 044028

\bibitem[{{Lousto} \& {Zlochower}(2009)}]{2009PhRvD..79f4018L}
---. 2009, \prd, 79, 064018

\bibitem[{{Lousto} {et~al.}(2012){Lousto}, {Zlochower}, {Dotti}, \&
  {Volonteri}}]{2012arXiv1201.1923L}
{Lousto}, C.~O., {Zlochower}, Y., {Dotti}, M., \& {Volonteri}, M. 2012, ArXiv
  e-prints

\bibitem[{{Lynden-Bell} \& {Eggleton}(1980)}]{1980MNRAS.191..483L}
{Lynden-Bell}, D. \& {Eggleton}, P.~P. 1980, \mnras, 191, 483

\bibitem[{{Mackey} {et~al.}(2008){Mackey}, {Wilkinson}, {Davies}, \&
  {Gilmore}}]{2008MNRAS.386...65M}
{Mackey}, A.~D., {Wilkinson}, M.~I., {Davies}, M.~B., \& {Gilmore}, G.~F. 2008,
  \mnras, 386, 65

\bibitem[{{Maraschi} {et~al.}(1976){Maraschi}, {Reina}, \&
  {Treves}}]{1976ApJ...206..295M}
{Maraschi}, L., {Reina}, C., \& {Treves}, A. 1976, \apj, 206, 295

\bibitem[{{McNamara} {et~al.}(2003){McNamara}, {Harrison}, \&
  {Anderson}}]{2003ApJ...595..187M}
{McNamara}, B.~J., {Harrison}, T.~E., \& {Anderson}, J. 2003, \apj, 595, 187

\bibitem[{{Merritt}(2009)}]{2009ApJ...694..959M}
{Merritt}, D. 2009, \apj, 694, 959

\bibitem[{{Merritt} {et~al.}(2001){Merritt}, {Ferrarese}, \&
  {Joseph}}]{2001Sci...293.1116M}
{Merritt}, D., {Ferrarese}, L., \& {Joseph}, C.~L. 2001, Science, 293, 1116

\bibitem[{{Milone} {et~al.}(2012){Milone}, {Piotto}, {Bedin}, {Aparicio},
  {Anderson}, {Sarajedini}, {Marino}, {Moretti}, {Davies}, {Chaboyer},
  {Dotter}, {Hempel}, {Mar{\'{\i}}n-Franch}, {Majewski}, {Paust}, {Reid},
  {Rosenberg}, \& {Siegel}}]{2012A&A...540A..16M}
{Milone}, A.~P., {Piotto}, G., {Bedin}, L.~R., {Aparicio}, A., {Anderson}, J.,
  {Sarajedini}, A., {Marino}, A.~F., {Moretti}, A., {Davies}, M.~B.,
  {Chaboyer}, B., {Dotter}, A., {Hempel}, M., {Mar{\'{\i}}n-Franch}, A.,
  {Majewski}, S., {Paust}, N.~E.~Q., {Reid}, I.~N., {Rosenberg}, A., \&
  {Siegel}, M. 2012, \aap, 540, A16

\bibitem[{{Ohsuga} {et~al.}(2005){Ohsuga}, {Mori}, {Nakamoto}, \&
  {Mineshige}}]{2005ApJ...628..368O}
{Ohsuga}, K., {Mori}, M., {Nakamoto}, T., \& {Mineshige}, S. 2005, \apj, 628,
  368

\bibitem[{{Peters}(1964)}]{1964PhRv..136.1224P}
{Peters}, P.~C. 1964, Physical Review, 136, 1224

\bibitem[{{Popham} {et~al.}(1999){Popham}, {Woosley}, \&
  {Fryer}}]{1999ApJ...518..356P}
{Popham}, R., {Woosley}, S.~E., \& {Fryer}, C. 1999, \apj, 518, 356

\bibitem[{{Popova} {et~al.}(1982){Popova}, {Tutukov}, \&
  {Yungelson}}]{1982Ap&SS..88...55P}
{Popova}, E.~I., {Tutukov}, A.~V., \& {Yungelson}, L.~R. 1982, \apss, 88, 55

\bibitem[{{Portegies Zwart} \& {McMillan}(2002)}]{2002ApJ...576..899P}
{Portegies Zwart}, S.~F. \& {McMillan}, S.~L.~W. 2002, \apj, 576, 899

\bibitem[{{Quinlan} \& {Shapiro}(1989)}]{1989ApJ...343..725Q}
{Quinlan}, G.~D. \& {Shapiro}, S.~L. 1989, \apj, 343, 725

\bibitem[{{Rezzolla} {et~al.}(2008){Rezzolla}, {Barausse}, {Dorband},
  {Pollney}, {Reisswig}, {Seiler}, \& {Husa}}]{2008PhRvD..78d4002R}
{Rezzolla}, L., {Barausse}, E., {Dorband}, E.~N., {Pollney}, D., {Reisswig},
  C., {Seiler}, J., \& {Husa}, S. 2008, \prd, 78, 044002

\bibitem[{{Rubenstein} \& {Bailyn}(1997)}]{1997ApJ...474..701R}
{Rubenstein}, E.~P. \& {Bailyn}, C.~D. 1997, \apj, 474, 701

\bibitem[{{Spitzer}(1987)}]{1987degc.book.....S}
{Spitzer}, L. 1987, {Dynamical evolution of globular clusters}, ed. {Spitzer,
  L.}

\bibitem[{{Strader} {et~al.}(2012){Strader}, {Chomiuk}, {Maccarone},
  {Miller-Jones}, {Seth}, {Heinke}, \& {Sivakoff}}]{2012ApJ...750L..27S}
{Strader}, J., {Chomiuk}, L., {Maccarone}, T.~J., {Miller-Jones}, J.~C.~A.,
  {Seth}, A.~C., {Heinke}, C.~O., \& {Sivakoff}, G.~R. 2012, \apjl, 750, L27

\bibitem[{{Strubbe} \& {Quataert}(2009)}]{2009MNRAS.400.2070S}
{Strubbe}, L.~E. \& {Quataert}, E. 2009, \mnras, 400, 2070

\bibitem[{{van Meter} {et~al.}(2010){van Meter}, {Miller}, {Baker}, {Boggs}, \&
  {Kelly}}]{2010ApJ...719.1427V}
{van Meter}, J.~R., {Miller}, M.~C., {Baker}, J.~G., {Boggs}, W.~D., \&
  {Kelly}, B.~J. 2010, \apj, 719, 1427

\end{thebibliography}

\end{document}